\documentstyle[11pt,aaspp4]{article}
\def\gax{\mathrel{\raise.3ex\hbox{$>$}\mkern-14mu\lower0.6ex\hbox{$\sim$}}}
\def\lax{\mathrel{\raise.3ex\hbox{$<$}\mkern-14mu\lower0.6ex\hbox{$\sim$}}}
\def\gtorder{\mathrel{\raise.3ex\hbox{$>$}\mkern-14mu
             \lower0.6ex\hbox{$\sim$}}}
\def\ltorder{\mathrel{\raise.3ex\hbox{$<$}\mkern-14mu
             \lower0.6ex\hbox{$\sim$}}}

\input psfig

\begin{document}

\title{Gravitational Lenses, the Distance Ladder and the Hubble Constant: \\
  A New Dark Matter Problem}

\author{C.S. Kochanek}
\affil{Harvard-Smithsonian Center for Astrophysics, 
       60 Garden Street, Cambridge, MA 02138}
\affil{email: ckochanek@cfa.harvard.edu}

\begin{abstract}
In cold dark matter models, a galaxy's dark matter halo is more spatially
extended than its stars.  However, even though the five well-constrained
gravitational lenses with time delay measurements must have similar dark 
matter distributions, reconciling the Hubble constant estimated from their 
time delays with local estimates is possible only if that dark matter 
distribution is as compact as the luminous galaxy.  The Hubble constant is 
$H_0=48_{-4}^{+7}$~km/s~Mpc (95\% confidence) if the lenses have flat rotation curves 
and $H_0=71\pm6$~km/s~Mpc (95\% confidence) if they have constant mass-to-light 
ratios, as compared to $H_0=72\pm8$~km/s~Mpc (68\% confidence) for local 
estimates by the HST Key Project.  Either all five $H_0$ estimates based on 
the lenses are wrong, local estimates of $H_0$ are too high, or dark matter 
distributions are more concentrated than expected.  The average value for
$H_0$ including the uncertainties in the mass distribution, $H_0=62\pm7$~km/s~Mpc, 
has uncertainties that are competitive with local estimates.  However, 
by selecting a value for $H_0$ you also determine the dark matter distribution 
of galaxies.
\end{abstract}

\keywords{cosmology: gravitational lensing; cosmology: Hubble constant; dark matter}

\section{Introduction}

Gravitational lens time delay measurements can determine the Hubble constant
$H_0$ given a model for the gravitational potential of the lens galaxy
(Refsdal~\cite{Refsdal64}).  The number of accurate delay measurements has begun 
to grow rapidly (see Schechter~\cite{Schechter00}), leading to a sample
of nine lenses.  Five of these systems have time delay measurements, relatively 
simple environments, and a dominant lens galaxy with regular isophotes 
(RXJ0911+0551, Burud~\cite{Burud02b}; PG1115+080, Schechter et al.~\cite{Schechter97}, 
Barkana~\cite{Barkana97}, Impey et al.~\cite{Impey98}; SBS1520+530, 
Burud~\cite{Burud02b}, Faure et al.~\cite{Faure02}; B1600+434, 
Burud et al.~\cite{Burud00}, Koopmans et al.~\cite{Koopmans00}; and 
HE2149--2745, Burud et al.~\cite{Burud02}).  
The remaining four systems either have poorly 
determined properties (the lens positions in B0218+357 and PKS1830--211, see
Lehar et al.~\cite{Lehar00}, Winn et al.~\cite{Winn02}, Courbin et al.~\cite{Courbin02}),
or more complicated lenses (interacting galaxies in B1608+656, 
Koopmans \& Fassnacht~\cite{Koopmans99}, a brightest cluster
galaxy in Q0957+561, see Keeton et al.~\cite{Keeton00}).  With a sample
of five relatively clean systems, we have reached the point where gravitational
lenses become a serious alternative to the local distance ladder for the
determination of $H_0$.

The determination of $H_0$ from a gravitational lens time delay requires a
model for the mass distribution of the lens galaxy.  In both models of particular
time delay lenses (e.g. Keeton \& Kochanek~\cite{Keeton97}, Impey et al.~\cite{Impey98},
Koopmans \& Fassnacht~\cite{Koopmans99}, Lehar et al.~\cite{Lehar00}, 
Keeton et al.~\cite{Keeton00}, Winn et al.~\cite{Winn02}) and from general 
analytic principles (Refsdal \& Surdej~\cite{Refsdal94}, 
Witt, Mao \& Schechter~\cite{Witt95}, Witt, Mao \& Keeton~\cite{Witt00},
Rusin~\cite{Rusin00}, Zhao \& Pronk~\cite{Zhao01}, Wucknitz~\cite{Wucknitz02}, Oguri et al.~\cite{Oguri02})
we know there is usually a degeneracy between the radial mass distribution
in the lens and the estimate of $H_0$, in the sense that more compact
mass distributions lead to larger Hubble constants for a fixed time
delay.  This degeneracy can be broken by finding additional constraints
(e.g. host galaxies, Kochanek, Keeton \& McLeod~\cite{Kochanek01}),
or by requiring the mass distribution to
be consistent with that of other lenses (e.g. Munoz, Kochanek \& Keeton~\cite{Munoz01}), 
weak lensing studies (e.g. Guzik \& Seljak~\cite{Guzik02}),
local estimates from stellar dynamical observations (e.g. Rix et al. 1997,
Romanowsky \& Kochanek~\cite{Romanowsky99}, Gerhard et al.~\cite{Gerhard01}, 
Treu \& Koopmans~\cite{Treu02}), X-ray observations (e.g. Fabbiano~\cite{Fabbiano89}, 
Lowenstein \& White~\cite{Lowenstein99}) or theoretical models (e.g. 
Kochanek \& White~\cite{Kochanek01b}, Keeton~\cite{Keeton01b}).  All these
direct estimates support a dark matter dominated model in which the lenses have flat
or slowly declining rotation curves near the Einstein ring.

We can also reverse the problem and simply determine the mass distribution the lenses
must have in order to agree with local estimates of $H_0$ (e.g. Rusin~\cite{Rusin00}).  
We will use the measurement by the HST Key Project (Freedman et al.~\cite{Freedman01}) 
of $H_0=72\pm8$~km/s~Mpc as our fiducial, independent estimate for $H_0$.  There are
two caveats about using this constraint.  First, the Key Project estimate is
significantly higher than the estimate of $H_0=59\pm6$~km/s~Mpc
by Saha et al.~(\cite{Saha01}) based on their Cepheid calibration for a sample of 
galaxies with Type Ia supernovae.  Second, the uncertainties in both estimates are dominated
by non-Gaussian systematic errors such as photometric calibrations, the distance to the 
LMC, Cepheid metallicity corrections and blending rather than statistical uncertainties    
(see the discussion in Freedman et al.~\cite{Freedman01}).

Even if the individual time delay lenses provide no direct constraints on their dark
matter distributions, our general understanding of galaxies and the CDM paradigm set
some broad limits.  In our analysis we consider two limiting possibilities for the 
mass distributions of the gravitational lenses.
The first model is a dark matter model with a flat rotation curve.
These models set a lower bound on $H_0$ unless typical galaxies
have rising rotation curves on the spatial scales of Einstein rings (typically
1.0--1.5$R_e$).  The second model is a constant mass-to-light ratio (constant $M/L$) 
model based on the photometry
obtained by the CfA/Arizona Space Telescope Lens Survey (CASTLES, Falco et al.~\cite{Falco01}).
No galaxy models in a cold dark matter (CDM) dominated universe should have mass distributions
which are more centrally concentrated than the luminosity distribution 
(e.g. Mo, Mao \& White~\cite{Mo98}).     
Hence, the constant $M/L$ models set an upper bound on $H_0$ (Impey et al.~\cite{Impey98}). 
In \S2 we discuss the data and our analysis methods.  In \S3 we present the estimates of
$H_0$ for the two limiting mass distributions and compare them to the local estimates.
We are forced to conclude that there is a problem in either the lens estimates of $H_0$, the
local estimates of $H_0$ or our understanding of dark matter distributions.

\section{The Available Data}

There are nine lenses with reasonably accurate time delay measurements 
(see Schechter~\cite{Schechter00}, Burud~\cite{Burud02b}) of which we consider only five: 
RXJ0911+0551, PG1115+080, SBS1520+530, B1600+434 and HE2149--2745. 
 The four lenses we do not include 
in our analysis are B0218+357, Q0957+561, B1608+656 and PKS1830--211.  
We first discuss why we used only five of the nine lenses. Then we provide a 
brief synopsis of the data and previous models of the five lenses we include 
in our analysis.  

In B0218+357 and PKS1830--211 the problem lies in making an unambiguous
measurement of the lens galaxy position relative to the images because
there is a strong degeneracy between the lens position and the value of $H_0$ 
even for a fixed mass distribution (see Lehar et al.~\cite{Lehar00}).
In B0218+357 the small image separation ($0\farcs35$) made it impossible
to reliably determine the lens position from existing HST images
(Lehar et al.~\cite{Lehar00}), with the best estimates favoring very
low values of $H_0$.  Using the centroid of the radio ring leads to a higher
value of $H_0$ (Biggs et al.~\cite{Biggs99}), but the center of an Einstein 
ring is more closely related to the position of the source than of the lens 
(see Kochanek et al.~\cite{Kochanek01}).  Wucknitz~(\cite{Wucknitz01}) 
estimated the lens position by modeling the radio ring. At this position
the delay implies very high values of $H_0$, but the models are inconsistent
with the VLBI data for the radio cores. Given these problems,
we do not include B0218+357 in our analysis.  A new attempt to
measure the position directly is planned using the Advanced Camera
for Surveys (ACS, Cycle 11, PI N. Jackson). 

While we are confident of our estimate of the lens position in PKS1830--211
(Winn et al.~\cite{Winn02}), an independent analysis of the same data
by Courbin et al.~(\cite{Courbin02}) suggests an alternate interpretation
involving multiple lens components.  Winn et al.~(\cite{Winn02}) find that
the lens is a single spiral galaxy with a well-determined position leading
to estimates of $H_0$ very similar to those for the five lenses we will analyze 
here ($H_0=44\pm9$~km/s~Mpc for an SIE model of PKS1830--211).  
The quality of the 
images makes it impossible to derive an accurate photometric model for the
galaxy, so we cannot produce an estimate of $H_0$ under the assumption of
a constant mass-to-light ratio. If, on the other hand, the multiple lens 
interpretation of Courbin et al.~(\cite{Courbin02}) is correct, then the 
PKS1830--211 system is useless for estimates of $H_0$ given the available
data.  Deeper HST images of 
the system could resolve the problem, but none are currently scheduled. 

We neglect Q0957+561 and B1608+656 because they are compound lenses with
many more parameters than constraints.  The Q0957+561 lens is dominated
by a brightest cluster galaxy sitting near the center of its cluster
(see Chartas et al.~\cite{Chartas02}).  Keeton et al.~(\cite{Keeton00},
also Bernstein et al.~\cite{Bernstein97})
were able to use the properties of the lensed images of the quasar
host galaxy to show that all earlier models of the system were 
inconsistent with the observed structure of the host galaxy, but
were unable to use the host galaxy to obtain a robust estimate
of $H_0$.  The structure of the host requires a cluster near the
lens galaxy, as confirmed by the Chartas et al.~(\cite{Chartas02})
X-ray observations, and it also requires that the lens galaxy dominate 
the image splitting.  Essentially all earlier estimates of $H_0$
using Q0957+561 violated these two requirements (see Keeton et al.~\cite{Keeton00}).
Very deep infrared images planned for HST Cycle 11 should provide the constraints 
needed to make reliable estimates of $H_0$ from this system.

In B1608+656 the lens consists of two interacting galaxies lying inside the 
Einstein ring defined by the images (see Koopmans \& Fassnacht~\cite{Koopmans99}).  
The two galaxies have irregular isophotes, due to the
interactions and the presence of dust, and the position estimates for the two
galaxies depend on the wavelength of the observation (see Koopmans \& 
Fassnacht~\cite{Koopmans99}, Surpi \& Blandford~\cite{Surpi01}).  Dark matter models
by Koopmans \& Fassnacht~(\cite{Koopmans99}) find higher values of $H_0$ than
for the lenses we analyze, with $H_0=66\pm7$~km/s~Mpc for their standard
model, although they obtain a broader range of values under different assumptions
for the position of the two lens galaxies and the degree to which the
model should be constrained by the observed orientations and ellipticities.
If the luminosity distributions are complex, it seems reasonable to expect
complex mass distributions which may not be well described by standard
mass models.  This system may be better modeled using the non-parametric
approach of Williams \& Saha~(\cite{Williams00}), which can include the
effects of transient mass distributions produced by the merger. For the
more regular lenses, the non-parametric solutions must be employed with
care because the positive surface density constraint used to limit
parameter space allows too much freedom in the models (permitting
negative density distributions, negative distribution functions or 
dynamically unstable solutions). This is in contrast to standard parametric 
models, which are guaranteed to correspond to physical dynamical models
at the price of restricting the freedom in the mass distribution.
While we regard our reasons for neglecting B1608+656
as legitimate, it is the one system we drop whose properties may run
counter to our general discussion.

We now discuss the five systems (RXJ0911+0551, PG1115+080, SBS1520+534,
B1600+434 and HE2149--2745) we include in our analysis.

We fit RXJ0911+0551 based on the CASTLES photometric data, the Burud~(\cite{Burud02b})
delay of $150\pm6$~days (68\% confidence) between the A-C cusp images and image D, and the 
Morgan et al.~(\cite{Morgan01}) centroid for the nearby X-ray cluster 
$12\farcs0\pm3\farcs0$ East and $-39\farcs8\pm3\farcs0$ South of image B.  
The lens galaxies were modeled by two de Vaucouleurs components.  The primary lens has a
major axis effective radius of $0\farcs77\pm0\farcs07$, a major axis
position angle of $-39^\circ\pm15^\circ$ and an axis ratio of $0.79\pm0.06$.
The satellite galaxy (only 8\% of the H-band flux) has a major axis effective 
radius of $0\farcs36\pm0\farcs14$, and is indistinguishable from round in
the available data.  To simplify the fits we used a round satellite in 
both the constant $M/L$ and dark matter lens models.  Test runs showed
this assumption had no significant effects on the estimates of $H_0$.
The isophotes of the two components are well separated and 
regular (unlike B1608+656).

The gravitational potential of the cluster near RXJ0911+0551 has significant
effects on the estimates of $H_0$.  We modeled the cluster as a singular
isothermal sphere (SIS), finding reasonable fits for cluster velocity
dispersions of $1100$~km/s for the dark matter models and $1250$~km/s 
for the constant $M/L$ models. The SIS models are in reasonable agreement
with the $840\pm200$~km/s velocity dispersion of the galaxies (Kneib, Cohen 
\& Hjorth~\cite{Kneib00}, see Burud~\cite{Burud02b}). An SIS model has
equal shear and convergence, $\kappa_c=\gamma_c=b_c/2r\simeq 0.25$ 
where $b_c$ is the cluster critical radius and $r$ is the distance to
the cluster center.  We could obtain Hubble constants $1/(1-\kappa_c)\simeq 0.38$ 
higher than those for the SIS models by assuming a very compact cluster
with $\kappa_c=0$ at the location of the lens because of the  
mass sheet degeneracy in the models (Falco, Gorenstein \& Shapiro~\cite{Falco85}).
Thus, our RXJ0911+0551 models really supply the likelihood for
$H_0(1-\kappa_c)/(1-\kappa_{SIS})$ where $\kappa_{SIS}$ is the
convergence for the SIS model and $\kappa_c$ is the convergence
for the true cluster density profile, both measured at the lens.  
Like the lens galaxies, the cluster rotation curve should also be
flat or falling on these scales and must have positive density,
so $0 \leq \kappa_c \leq \kappa_{SIS}$.  This means that we can use
RXJ0911+0551 only to obtain lower bounds on $H_0$ or to estimate
the true cluster surface density $\kappa_c$. 

We fit PG1115+080 based on the photometric data from Impey et al.~(\cite{Impey98}) and
the time delay estimates by Barkana~(\cite{Barkana97}, based on
Schechter et al.~\cite{Schechter97}). The longest delay is $25\pm4$~days (95\% confidence).  
The model consists of a de Vaucouleurs
primary lens galaxy and a perturbing isothermal group.  Kochanek et al.~(\cite{Kochanek01})
found that the structure of the Einstein ring image of the quasar host galaxy
weakly (2$\sigma$) ruled out constant $M/L$ models in favor of models with
a flat rotation curve and lower values of $H_0$.  Historically, the low   
estimates of $H_0$ from PG1115+080 have been regarded as anomalous (e.g.
see Koopmans \& Fassnacht~\cite{Koopmans99}).  Our models reproduce the
earlier estimates by Impey et al.~(\cite{Impey98}).

We fit SBS1520+530 based on the CASTLES photometric data (see also Faure et al.~\cite{Faure02})
and the Burud~(\cite{Burud02b}) delay of $130\pm3$~days (68\% confidence). The lens was constrained
to be a de Vaucouleurs model with a major axis effective radius of $R_e=0\farcs60\pm0\farcs02$,
a major axis position angle of $-26^\circ\pm2^\circ$ and an axis ratio of $0.46\pm0.04$.
The models included an external shear restricted by a Gaussian prior to the
range $\gamma=0.05\pm0.05$ (see the discussion of HE2149--2745).

We fit B1600+434 based on photometric fits to the CASTLES data and the optical time 
delay of $51\pm4$~days (95\% confidence) measured by Burud et al.~(\cite{Burud00}).  
The optical delay agrees with 
the radio delay of $57\pm6$~days (68\% confidence) 
measured by Koopmans et al.~(\cite{Koopmans00}) but it has smaller
uncertainties.  In the constant $M/L$ models we did not constrain the mass ratio
of the disk and the bulge, which allowed the models more freedom than strictly 
necessary.  The constant $M/L$ model consisted of a de Vaucouleurs bulge and
an exponential disk forced to have a common centroid and major axis position angle 
($44^\circ\pm6^\circ$).  The bulge (disk) has a major axis scale length of 
$0\farcs45\pm0\farcs06$ ($0\farcs84\pm0\farcs06$) and an axis ratio
of $0.69\pm0.05$ ($0.16\pm0.06$). These values are broadly consistent with
the estimates by Maller et al.~(\cite{Maller00}), differing mainly in our
requirement that the two components share a common orientation.  The dust 
lane in the lens galaxy makes its position relatively uncertain, and we
adopt the position of ($\Delta\hbox{RA}=-0\farcs13\pm0\farcs05$,
$\Delta\hbox{Dec}=0\farcs21\pm0\farcs05$) from image B used by 
Koopmans, de Bruyn \& Jackson~(\cite{Koopmans98}).  The estimates for $H_0$ 
in Burud et al.~(\cite{Burud00}) and Koopmans et al.~(\cite{Koopmans00}) used 
hybrid (disk $+$ bulge $+$ halo) mass models developed by Koopmans et al.~(\cite{Koopmans98}) 
and Maller et al.~(\cite{Maller00}). These models give values of $H_0$ between our 
our dark matter and constant $M/L$ limiting models for the same lens position.

We fit HE2149--2745 based on the CASTLES photometric data and the time delay of
$103\pm12$~days (68\% confidence) reported
by Burud et al.~(\cite{Burud02}).  The model consists of a de Vaucouleurs
primary lens galaxy in a perturbing external (tidal) shear.  We must always allow 
for a local shear in realistic models 
(see Keeton, Kochanek \& Seljak~\cite{Keeton97}, Kochanek~\cite{Kochanek02}), 
although it has little effect on the estimate of $H_0$
($\delta H_0/H_0 \sim \gamma$, see Witt et al.~\cite{Witt00}). The external shear is 
constrained by a weak prior ($\gamma=0.05 \pm 0.05$) to the typical range observed in 
four-image lenses (see Keeton et al.~\cite{Keeton97b}, Kochanek~\cite{Kochanek02}).  
In RXJ0911+0551, PG1115+080 and B1600+434, the shear perturbations were supplied by
the nearby group or galaxy. 

Burud et al.~(\cite{Burud02}) modeled the
system using pseudo-Jaffe models (Keeton \& Kochanek~\cite{Keeton97},
Munoz et al.~\cite{Munoz01}) to explore the dependence of $H_0$ on the radial
mass distribution. They adopted a break radius for the density distribution such
that the ellipticities of the mass and the light agreed leading to an estimate
of $H_0=66\pm8$~km/s~Mpc.  This criterion is arbitrary because we only have 
evidence for the alignment of the mass and the light in lens galaxies 
rather than for a similarity in shape (see Kochanek~\cite{Kochanek02}).  Based on
theoretical arguments and other observations that halos tend to be rounder
than the luminous galaxies (see Sackett~\cite{Sackett99}), we can plausibly 
use the observed ellipticity only as an upper bound on the models.  Models of 
HE2149--2745 with more dark matter (a larger break radius) are rounder, just
as we might expect, and allow significantly lower values of $H_0$.  The
addition of the external shear further loosens any constraint associated
with the ellipticity of the lens.

We modeled these five lenses using the {\it lensmodel} package (Keeton~\cite{Keeton01}).
Each lens was fit using either a dark matter model or a constant $M/L$ model.
The dark matter model used a singular isothermal ellipsoid 
(SIE) to represent the overall mass distribution of the system.  The constant
$M/L$ models used ellipsoidal de Vaucouleurs models and exponential disks
constrained by the photometric model for the galaxies.  We use the photometry
only as a constraint on the mass distribution -- the masses of the 
components are determined by the lens models with no constraints based on 
estimates of stellar mass-to-light ratios.  
Perturbing galaxies and clusters
were modeled as singular isothermal spheres (SIS).  All models included 
either an SIS perturber (RXJ0991+0551, PG1115+080, B1600+434) or a constrained 
tidal shear (SBS1520+530, HE2149--2745).  
We assumed 20\% errors in the image fluxes,
where the uncertainties are dominated by systematic problems such as substructure 
(see Dalal \& Kochanek~\cite{Dalal02}) rather than measurement uncertainties.  
We also adopt a minimum time delay error of 5\% to encompass systematic errors 
such as convergence fluctuations from large scale structure (e.g. Seljak~\cite{Seljak94},
Barkana~\cite{Barkana96}).    The goodness of fit was determined by a $\chi^2$
statistic measured on the lens plane for the images combined with the fit
to the observed time delays.  Both the dark matter and the constant $M/L$
models produce statistically acceptable fits to the image configurations 
all lenses except RXJ0911+0551.  For this system the constant $M/L$ model
is formally ruled out ($\Delta\chi^2=20$), but this is in part an artifact
of using a round satellite galaxy to speed the calculations.  
We use maximum likelihood
methods to analyze the results by examining the likelihood ratio
$2 \ln (L/L_{max}) = \chi^2(H_0)-\chi^2_{min}(H_0)$ for each model 
sequence.   We adopt an $\Omega_0=0.3$ flat cosmological model for our analysis
and discussion.\footnote{All cited values of $H_0$ from the lenses are scaled
to this cosmological model.  While the $H_0$ estimates from lenses do depend
on the cosmological model (Refsdal~{\protect\cite{Refsdal66}}), the magnitude of the 
variations is too small to be
relevant to our discussion.  A demonstration that the cosmological model
estimated by lens time delay measurements agrees with other estimates will
be important as part of any claim that the errors in the $H_0$ measurements
are smaller than about 5\%.  As an experiment, we estimated $\Omega_m$ for 
flat cosmological models based on the estimates of $H_0$
for the dark matter models, finding that $\Omega_m=1.0$ is ruled out
compared to our standard model at $0.1\sigma$!}

\begin{figure}
\centerline{\psfig{figure=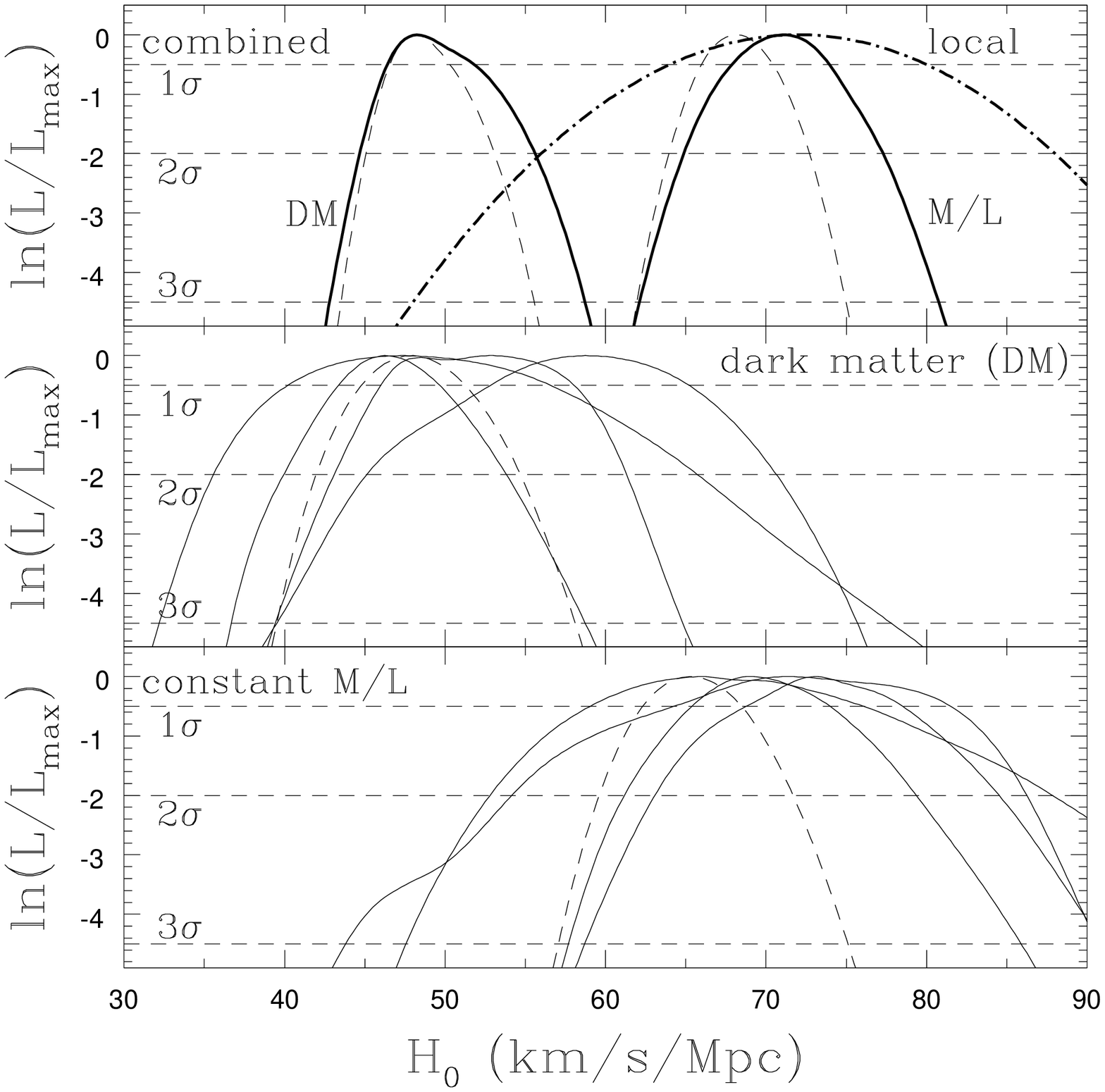,width=6.0in}}
\caption{ Maximum likelihood estimates for $H_0$.  \newline
  (Bottom) The likelihood functions for the
  constant $M/L$ models of the individual lenses.  Obtaining higher
  values for $H_0$ than the constant $M/L$ models requires dark 
  matter distributions which are more compact than the luminosity
  distribution. The dashed curve is RXJ0911+0551. \newline
  (Middle) The likelihood functions for the
  dark matter (DM) models of the individual lenses. 
  Obtaining lower values for $H_0$ than the DM models requires
  a mass distribution with a rising rotation curve at the 
  Einstein ring of the lens (at 1--2$R_e$). The dashed curve is
  RXJ0911+0551. \newline
  (Top) The upper panel shows the combined likelihood functions for
  the dark matter (DM) models (left solid/dash), the constant $M/L$
  models (right solid/dash) and a Gaussian model for the local 
  estimate by the HST Key project (dashed-dot).  The solid (dashed)
  curves for the dark matter and constant $M/L$ models exclude
  (include) RXJ0911+0551.  
  \newline
  The horizontal dashed lines show the 68\% ($1\sigma$), 95\%
  ($2\sigma$) and 99\% ($3\sigma$) confidence limits based on the 
  likelihood ratio. 
  \label{fig:results}
  }
\end{figure}

\begin{figure}
\centerline{\psfig{figure=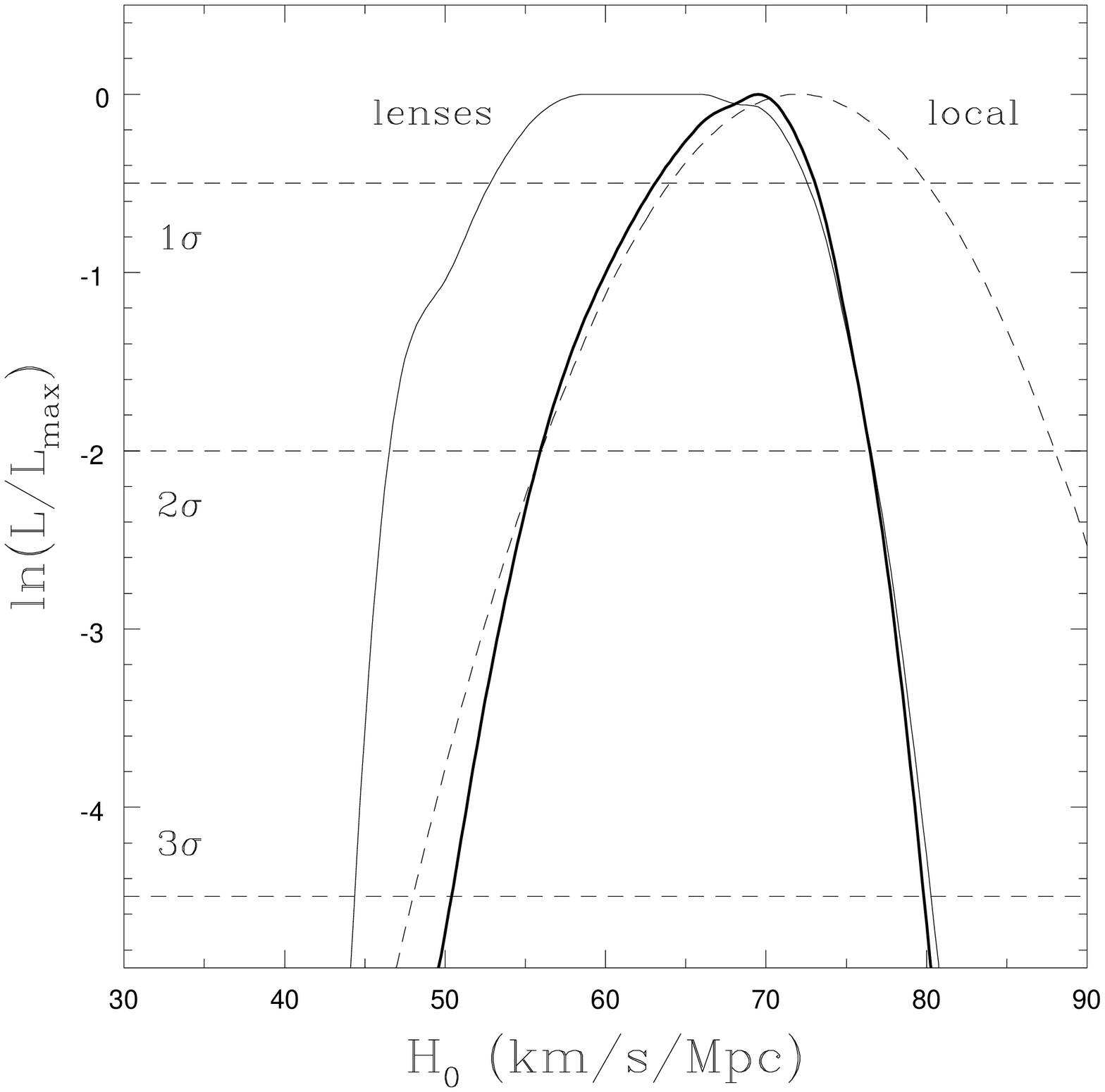,width=6.0in}}
\caption{ Final likelihood estimates for $H_0$. 
  The light solid line shows the estimate from the lens time delays assuming the
  mass distribution is bounded by the dark matter and constant $M/L$ models but
  we are unable to discriminate between any intermediate model.  The lower bounds
  are due to the dark matter models and the upper bounds are due to the 
  the constant $M/L$ models.  RXJ0911+0551 is included assuming the cluster
  surface density is bounded by $\kappa_{SIS} \geq \kappa_c \geq 0$. The dashed 
  line is the HST Key Project estimate and the heavy solid line is the 
  joint likelihood.   Compared to the mean and variance of $H_0=72\pm8$~km/s~Mpc
  of the Key Project, the lenses have $H_0=62\pm7$~km/s~Mpc and the joint
  likelihood has $H_0=67\pm5$~km/s~Mpc.  The variance in the $H_0$ estimate 
  differs from the 68\% confidence region of $53~\hbox{km/s~Mpc} < H_0 < 73~\hbox{km/s~Mpc}$
  because the likelihood distribution is not Gaussian.  
  The problem with the region permitted
  by the joint likelihood is that it corresponds to models with little dark
  matter compared to the expectations for CDM.
  \label{fig:results2}
  }
\end{figure}

\section{Results and Conclusions}

As discussed in \S1, we focused on the dark matter and constant $M/L$ models because they represent
the limiting cases for physically possible mass distributions given our current
understanding of dark matter.  Unless  galaxies have rising rotation curves
near the Einstein ring of the lens (typically 1.0--1.5 effective radii
from the lens galaxy), the SIE models provide lower bounds on $H_0$ for a 
fixed time delay.  
Constant $M/L$ models are the most compact mass
distributions allowed for a gravitational lens, so they lead to firm
upper bounds on the $H_0$ (see Impey et al.~\cite{Impey98}).  Dark matter
cannot be more centrally concentrated than the stars unless their are
fundamental flaws in our understanding of cold dark matter.
These are two extreme limits, and all our theoretical expectations and estimates of 
mass distributions in other lenses or by other means  suggest
that reality must be closer to the dark matter model than to the 
constant $M/L$ model.  Our results, in the form of model likelihoods as a function of the Hubble
constant, are shown in Figures~1 and 2.  

Under either assumption about the mass distribution, the five lenses produce 
consistent estimates for $H_0$ (see Figure 1).  If we exclude
RXJ0911+0551, because of the added uncertainty created by the dark matter 
distribution of the nearby cluster (see below), we find that $H_0=48_{-4}^{+7}$~km/s~Mpc 
(95\% confidence) for the dark matter model and $H_0=71\pm6$~km/s~Mpc 
(95\% confidence) for the constant $M/L$ model. The jackknife and bootstrap
error estimates are comparable to those from the combined likelihoods.
Because the $H_0$ estimates for the lenses are very similar when we assume they have
the same radial mass profiles, the radial mass distributions must be 
similar.  For example, for potentials $\phi \sim R^\beta$, where $\beta=1$ is the 
isothermal profile we use for our dark matter model and the limit 
$\beta\rightarrow 0$ is a point mass, estimates of the Hubble constant
roughly scale as $H_0 \propto (2-\beta)$ (Witt et al.~\cite{Witt00}).  
If we assume that three of the four lenses have the dark matter ($\beta=1$) 
profile, then we find that the slope for the remaining lens must be 
        $\beta=0.83_{-0.16}^{+0.18}$,     
              $1.09_{-0.21}^{+0.09}$,     
              $1.19_{-0.15}^{+0.11}$ and  
              $0.98_{-0.28}^{+0.22}$      
(68\% confidence)
for PG1115+080, SBS1520+530, B1600+434 and HE2149--2745 respectively.  The 
jackknife estimate for the variance in the slopes is $\sigma_\beta\simeq 0.23$,
roughly half of which is due to the measurement uncertainties. The 
intrinsic scatter must be relatively small ($\sigma_\beta \simeq 0.15$). 
As a result, our estimates of $H_0$ vary little if we use any three of the four 
lenses (which is unfortunate, see the Appendix).

Despite the presence of the cluster, RXJ0911+0551 agrees with this general
picture.  If we assume that the lens galaxy in RXJ0911+0551
has a similar structure to the other lens galaxies, then we can estimate
the dark matter surface density in the cluster from the requirement
that all five systems must agree on the value of $H_0$.  We find
that the cluster convergence is $\kappa_c=0.24\pm0.08$ for the dark
matter models and $\kappa_c=0.25\pm0.06$ for the constant $M/L$
models. These values are very similar to the SIS convergences
of $\kappa_{SIS}=0.25$ and $0.31$, as we might expect when the lens
lies $200h^{-1}$~kpc from the cluster center where the cluster density
profile is close to $\rho \sim r^{-2}$ for typical models
(e.g. Bullock et al.~\cite{Bullock01}).

Unfortunately, while the time delay lenses must have very similar mass 
profiles, we cannot determine which profile is appropriate given the 
available data.  We can try to break the degeneracy by using an independent 
estimate of $H_0$ (e.g. Rusin~\cite{Rusin00}), in particular the local estimate 
by the HST Key Project that $H_0=72\pm8$~km/s~Mpc (Freedman et al.~\cite{Freedman01}).  
The agreement between the lenses and the Key Project is very
good for the constant $M/L$ models and terrible for the dark
matter models (see Figure 1).  Even with the freedom to adjust the cluster it
is difficult to reconcile the dark matter model of RXJ0911+0551 
with the Key Project estimate because it requires a negative
cluster surface density ($\kappa_c=-0.13\pm0.20$).  The local
estimate breaks the degeneracy, but it also means that 
the lens galaxies cannot have massive, extended dark halos. Such
a conclusion disagrees with the CDM picture of galaxies and the abundant
evidence for halos in galaxies and gravitational lenses similar
to the time delay systems.

\noindent There are three possible solutions.  

First, we may fundamentally misunderstand the distribution of dark matter 
in galaxies.  Our problem is the opposite of the more familiar ``dark matter 
crisis'' where models of dwarf and low surface brightness galaxy rotation curves 
seem to require dark matter distributions which are less centrally concentrated 
than theoretical predictions (see the review by Moore~\cite{Moore01}).  
The issue for fitting rotation curves is the structure of central density cusps, 
with gravitational lenses 
favoring steeper cusps than those expected from CDM models rather than the 
shallower cusps suggested by the rotation curves (see Rusin \& Ma~\cite{Rusin01},
Keeton~\cite{Keeton01b}, 
Munoz et al.~\cite{Munoz01}).  Whatever the structure of the cusps, all these
models still have dark matter distributions which are less concentrated than
the stars (baryons), while we need to make the dark matter at least as concentrated
as the stars in order to solve the $H_0$ problem illustrated by Fig.~1.  
At present, the lensing constraints on the 5 systems cannot determine 
the radial mass distributions robustly.
However, the lenses where we can make such estimates favor models close to
the dark matter limit (see Munoz et al.~\cite{Munoz01} and references therein),
as do weak lensing studies (e.g. Guzik \& Seljak~\cite{Guzik02}),
local stellar dynamical observations (e.g. Rix et al.~\cite{Rix97},
Romanowsky \& Kochanek~\cite{Romanowsky99},
Gerhard et al.~\cite{Gerhard01}, Treu \& Koopmans~\cite{Treu02}), and X-ray observations (e.g. 
Fabbianno~\cite{Fabbiano89}, Lowenstein \& White~\cite{Lowenstein99}).

Second, the Key Project estimate that $H_0=72\pm8$~km/s~Mpc could be too high
or have significantly underestimated uncertainties.  For example, the 
lensing results are more compatible with the estimate of 
$H_0=59\pm6$~km/s~Mpc from Cepheid calibrations of the distances to
Type Ia supernovae (Saha et al.~\cite{Saha01}).  Freedman et al.~(\cite{Freedman01}) 
argue strongly against the Saha et al.~(\cite{Saha01}) analysis for
a range of legitimate technical issues, which is why we adopted
the Key Project estimate as our standard for local estimates.
There are also arguments for significantly larger corrections for metallicity
(e.g. Sasselov et al.~\cite{Sasselov97}, Kochanek~\cite{Kochanek97}) and blending (e.g. Mochejska
 et al.~\cite{Mochejska00}) in the Cepheid distances.  Finally, the 
standard error in the Key Project estimate of $H_0$ is dominated by 
a quadrature sum of systematic errors.  This makes a very specific 
assumption about the probability distributions of the systematic errors,
in the sense that conspiracies between the errors to systematically
shift the estimate of $H_0$ in one direction are assumed to be very unlikely. A
more conservative assumption about the behavior of systematic errors, such
as a direct sum of the systematic errors rather than a quadrature sum, would lead to significantly 
larger uncertainties and less of a conflict with the estimates from the lenses.  
On the other hand, the check of 
the Cepheid distance scale against the maser distance to NGC4258 suggests 
that the Key Project distances may moderately underestimate $H_0$ 
(Newman et al.~\cite{Newman01}).

The third possibility is that the estimates of $H_0$ from gravitational lenses
are systematically low.  Our result is a significant change from the
previous analysis of the available lens sample by Koopmans \& Fassnacht~(\cite{Koopmans99}),
who found $H_0=68\pm7$~km/s~Mpc for the dark matter models using
B0218+357, Q0957+561, PG1115+080, B1608+656 and PKS1830--211.  In our
analysis we used only one of these 5 systems and added the four new 
systems RXJ0911+0551, SBS1520+534, B1600+434 and HE2149--2745.  The high value 
found by Koopmans \& Fassnacht~(\cite{Koopmans99}) was due to including
the lenses we rejected as being unusable because the lens galaxy
position was unknown or the system is an under-constrained  
compound lens (see \S2 for a full discussion).  For example,
Koopmans \& Fassnacht~(\cite{Koopmans99}) assign an estimate of 
$H_0\simeq 85$~km/s~Mpc to PKS1830--211, while the best current 
estimate is $H_0=44\pm9$~km/s~Mpc due to an improved, but disputed,
direct estimate of the lens position (Winn et al.~\cite{Winn02},
Courbin et al.~\cite{Courbin02}).  The system with interacting
lens galaxies, B1608+656, is the one system which may disagree with our
present analysis.  

Given that the five lenses we use are relatively clean, well constrained,
and produce consistent results, it is unlikely that errors in the time
delay measurements or the model constraints are responsible for the 
results.  Nonetheless, the systems should be monitored further to both 
confirm the delay estimates and reduce the measurement errors to the point 
(under 5\%) where they make a negligible contribution to the overall 
uncertainties.  Deep, high resolution imaging to measure the structure of 
the lensed host galaxies will provide additional constraints on the lens models 
and can be used to determine the radial density profile (Kochanek et al.~\cite{Kochanek01}).
Lensed images of the host galaxy are seen in four of the current lenses
with time delays (Q0957+561, Keeton et al.~\cite{Keeton01};
PG1115+080, Impey et al.~\cite{Impey98}; B1600+434, Kochanek et al.~\cite{Kochanek99};
and B1608+656, Koopmans \& Fassnacht~\cite{Koopmans99}).  Additional
HST observations are the key to including the rejected systems in 
later analyses. The extra constraints from deep images of the host galaxies 
in Q0957+561 and B1608+656 may allow us to include them despite being compound
lenses, and improved HST imaging of B0218+357 and PKS1830--211 can resolve
the problems associated with the lens galaxy positions. 
The most important check on the results from the lenses is to measure precise
time delays in more systems, particularly since the statistics of time delays
provides another probe of the mass distribution (see Oguri et al.~\cite{Oguri02}).
  In order to avoid the current situation, in which
4 of 9 lenses with well-measured time delays are difficult to include in
quantitative estimates of $H_0$, monitoring campaigns should focus on systems
with one dominant lens galaxy having regular isophotes whose properties can be
easily measured with HST observations.

We considered only the extreme limits for the radial mass distribution of the
lenses, and we could obtain intermediate values of $H_0$ by slowly adding 
a dark matter halo to the constant $M/L$ models.  
It is easy to mimic such models, and the results are
shown in Figure~2.\footnote{For each lens, we set the likelihood ratio
to be that of the dark matter (constant $M/L$) model for $H_0$ below (above)
the Hubble constant corresponding to the model's maximum likelihood.
Since we are assuming no ability to distinguish the best fitting
constant $M/L$ and dark matter models, the likelihood ratios for all
intermediate values of $H_0$ are set to unity. Thus, the likelihood
distribution for each lens is a top hat with edges set by the 
physically limiting mass models softened by the uncertainties in
the model fits. }  Despite including the full, physically plausible
range of uncertainties in the dark matter distributions, the uncertainties
in the $H_0$ estimates from the lenses are comparable to those of
the local distance scale.  The mean and variance of the $H_0$ estimate
from the lenses is $H_0=62\pm7$~km/s~Mpc, compared to $72\pm8$ for the
Key Project.  The shapes of the distributions differ in detail because
the systematic errors for the lenses were modeled as a top hat while 
those of the Key Project were modeled as Gaussians.  Combining the two
likelihood distributions gives $H_0=67\pm5$~km/s~Mpc, which has reduced 
uncertainties but implies lens galaxies with little dark matter. 
Whatever the final resolution to the apparent conflicts, gravitational 
lenses are rapidly approaching the precision of the local distance scale 
methods of determining the Hubble constant despite the uncertainties 
and degeneracies in the present generation of models.

\noindent Acknowledgments.  I thank I. Burud, L. Koopmans, D. Rusin, D. Sasselov, 
P. Schechter, K. Stanek, J. Winn and S. Wyithe for their comments.  Most of the
time delays come from I. Burud's marvelous thesis, where she demonstrated that 
the mass production of time delays was a feasible project.  CSK is 
supported by the Smithsonian Institution and NASA ATP grants NAG5-8831 and NAG5-9265.  

\appendix

\section{Determining $H_0$ From Lens Samples With Inhomogeneous Mass Distributions}

When we started this analysis, the motivation was to explore the consequences for 
the determination of $H_0$ from using time delay lenses having a range of radial
mass distributions.   This seemed a requirement since available summaries 
(e.g. Koopmans \& Fassnacht~\cite{Koopmans99}) favored giving PG11115+080 
a tidally truncated halo so as to bring it into agreement with the remainder
of the sample. However, once we defined the requirements for our clean lens sample, 
the need for dissimilar mass distributions evaporated.   Far from being useful,
this convergence of the mass distributions makes the problem harder because a sample 
of lenses with time delays and very different dark matter distributions can be used 
to estimate $H_0$ even though every individual lens suffers from a model degeneracy 
between the estimate of $H_0$ and the dark matter distribution.

The basic argument is easily understood using the simple analytic scalings
of Witt et al.~(\cite{Witt00}) where there is a degeneracy between $H_0$
and the exponent of the density distribution, $\Delta t \propto (2-\beta)/H_0$
for $\phi \propto R^\beta$ and $0 \leq \beta \leq 1$ covers the range from
point masses to flat rotation curves.  Suppose we have a sample of lenses
whose true density exponents are $\beta_{true,i}$.  Given the true Hubble
constant $H_{true}$, we measure time delays 
$\Delta t_i \propto (2-\beta_{true,i})/H_{true}$.
We then model the systems assuming a total degeneracy in the lens models,
finding an estimates for the Hubble constant of 
$H_i=H_{true}(2-\beta)/(2-\beta_{true,i})$ for a model exponent of $\beta$.  
When all lenses have similar
mass distributions, $\beta_{true,i} \simeq \beta_{true,j}$, then the
Hubble constant estimates suffer from the global degeneracy we observe
in the current lens data -- the requirement that all the lenses agree
on the same value for the Hubble constant provides little leverage for
separating the effects of $\beta$ and $H_0$ on the time delay.  

If, however, the true mass distributions are widely scattered, then the simple
requirement that all lenses must agree on the same value of $H_0$ allows us to
determine $H_0$ even though every individual lens suffers from a complete
degeneracy between $H_0$ and $\beta$ restricted only by the minimal
requirement that the distribution is bounded by the limits of a flat rotation
curve ($\beta=1$) and a point mass ($\beta \rightarrow 0$).  Each lens
permits values of $H_0$ bounded by 
$H_{true}/(2-\beta_{true,i})\leq H_0 \leq 2H_{true}/(2-\beta_{true,i})$
where the lower (upper) limit corresponds to a model with $\beta=1$ ($\beta=0$).
If $0 \leq \beta_{min} \leq \beta_{true,i} < \beta_{max} \leq 1$, then
the ensemble of lenses restricts the Hubble constant to the
range $H_{true}/(2-\beta_{max}) \leq H_0 \leq 2H_{true}/(2-\beta_{min})$.  
The fractional range of the estimates,
\begin{equation}
  { \Delta H \over H_{true}} = { 1 \over 2-\beta_{min}}
   \left[ 1 - { \beta_{max}-\beta_{min} \over 2 - \beta_{max} } \right],
\end{equation}
is always reduced by having a spread of mass distributions, becoming
zero when the range of the true distributions matches that imposed on
the models.  Thus, it is better to have a mixture of lenses with and
without dark matter than to have the homogeneous mass distributions 
the data seem to require.

\end{document}